\documentclass[12pt]{iopart}
\usepackage{amssymb,epsfig,float}
\begin{document}
\title{ Ellipsoidal universe in the brane world}
\author{Xian-Hui Ge}
\address{Asia Pacific Center for Theoretical Physics, Pohang
790-784, Korea} \ead{gexh@apctp.org}
\author{Sang Pyo Kim}
\address{Department of Physics, Kunsan National University,
Kunsan 573-701, Korea\\Asia Pacific Center for Theoretical Physics,
Pohang 790-784, Korea}\ead{sangkim@kunsan.ac.kr}
\date{\today}
\begin{abstract}
We study a scenario of the ellipsoidal universe in the brane world
cosmology with a cosmological constant in the bulk . From the
five-dimensional Einstein equations we derive the evolution
equations for the eccentricity and the scale factor of the universe,
which are coupled to each other. It is found that if the anisotropy
of our universe is originated from a uniform magnetic field inside
the brane, the eccentricity decays faster in the bulk in comparison
with a four-dimensional ellipsoidal universe. We also investigate
the ellipsoidal universe in the brane-induced gravity and find the
evolution equation for the eccentricity which has a contribution
determined by the four- and five-dimensional Newton's constants. The
role of the eccentricity is discussed in explaining the quadrupole
problem of the cosmic microwave background.
\end{abstract}

\section{Introduction}

The WMAP three-year results show that the CMB anisotropy data are in
a remarkable agreement with the simplest inflation model, but
interestingly the large-scale feature still warrant further
attention \cite{Spergel}. The suppression of power spectrum at large
angular scales ($\theta\geq 60^{\circ}$), which is reflected in the
most distinguishable way in the reduction of the quadrupole
$\mathcal C_{2}$, remains unexplained by the standard inflation
model. Several authors suggest that the low multipole anomalies in
the CMB fluctuations maybe a signal of a nontrivial cosmic topology
\cite {col,Bunn,cor}. More precise measurements of WMAP showed that
the quadrupole $C_{2}$ and octupole $C_{3}$ are unusually aligned
and are concentrated in a plane inclined about $30^{\circ}$ to the
Galactic plane \cite{wmap}. This motivated an asymmetric expansion
universe model, in which one direction expands differently from the
other two (transverse) directions of the equatorial plane
\cite{bere}. It was further found that if the large-scale spatial
geometry of our universe is plane symmetric with an eccentricity at
decoupling of order $10^{-2}$, the quadrupole amplitude can be
drastically reduced without affecting higher multipoles of the
angular power spectrum of the temperature anisotropy
\cite{campanelli}.

In this paper, we explore a scenario of the ellipsoidal
 universe in the brane
 world. The brane cosmology of a 3-brane universe in a
five dimensional spacetime has been investigated in Refs.
\cite{deff,def,bin}. The Friedmann equation for such a brane
cosmology shows that the square of the Hubble parameter $H$ depends
quadratically on the brane energy density, whenever it depends
linearly on the matter energy density in the standard cosmology
\cite{bin}. We now extend the isotropic 4D spacetime to an
anisotropic spacetime in the brane cosmology, find the equations for
the isotropic scale factor as well as the eccentricity for
anisotropy and then relate the eccentricity with the quadrupole
anisotropy of CMB.

The organization of this paper is as follows. In Sec. II, we derive
the basic equations of a 4D ellipsoidal universe and find that for
the isotropic part of the total energy-momentum tensor, the
evolution of energy density is described by $\rho^{I}\propto
(1-e^2)^{-(\omega+1)}a^{-3(w+1)}$, where $e$ is the eccentricity of
the universe and $\omega$ is a parameter for the equation of state.
In Sec. III, we find the 5D Einstein equations for an ellipsoidal
universe with a cosmological constant in the bulk. And we obtain the
equations governing the evolution of the scale factor and the
eccentricity and then discuss their cosmological consequences. In
particular, we show that if anisotropy of the our universe is mainly
originated from a uniform magnetic field inside the brane, the
eccentricity decays faster in the bulk. In Sec. IV, we present the
evolution of the eccentricity in the brane-induced gravity and find
some higher order contributions due to the presence of an extra
dimension. In Sec. V, we discuss the relation between the
eccentricity in the brane and the CMB anisotropy and conclude the
paper.

\section{Four dimensional planar symmetric universe}

In this section, we derive the basic equations that describe the
evolution of a 4D ellipsoidal universe. The general plane-symmetric
metric is given by \cite{Taub}
\begin{equation}
\label{metric} ds^2 = dt^2 - a^2(t) (dx_{1}^2 + dx_{2}^2) - c^2(t)
\, dx_{3}^2,
\end{equation}
where the scale factors $a$ and $c$ are functions of the cosmic time
$t$ only and $x_{1}x_{2}$ denote the coordinates of the plane of
symmetry. In terms of the eccentricity defined by
\begin{equation}
\label{obla}
e=\sqrt{1-\left(\frac{c}{a}\right)^{2}},
\end{equation}
the metric can be rewritten as
\begin{equation}
ds^2 = dt^2 - a^2(t) (dx_{1}^2 + dx_{2}^2)-(1-e^{2}(t)) a^2(t)
dx_{3}^2.
\end{equation} Notice that we have tentatively assumed that $a\geq
c$. In fact, whether the shape of the universe is an oblate ($a\geq
c$) or prolate ($a\leq c$) spheroid depends on the form of
anisotropic energy-momentum density. The eccentricity for a prolate
sphere ($a\leq c$) is given by another form, $
e=\sqrt{1-\left(\frac{a}{c}\right)^{2}}$. Here, we should justify
the use of an oblate spheroid and the corresponding eccentricity of
the form (\ref{obla}).
 As we have mentioned the universe would have expanded
isotropically before an anisotropic expansion would have become
important. The isotropic tension densities plus anisotropic tension
densities would have caused the spherically symmetric sphere evolve
into a spheroid. At a later stage of the evolution of the universe,
when all of the contributions except the vacuum energy (cosmological
constant) faded away, the longitudinal and transverse directions
expand in an equal ratio and the expansion became isotropic. It was
proved in Ref. \cite{bere} that if the isotropy symmetry of the
universe is broken by a uniform magnetic field or a cosmic string,
then the resulting shape of the universe is an oblate sphere (see
figure~1). In our discussions below, we focus mainly on a homogenous
but anisotropic universe, where the anisotropy is contributed by a
uniform magnetic field.
\begin{figure}
\psfig{file=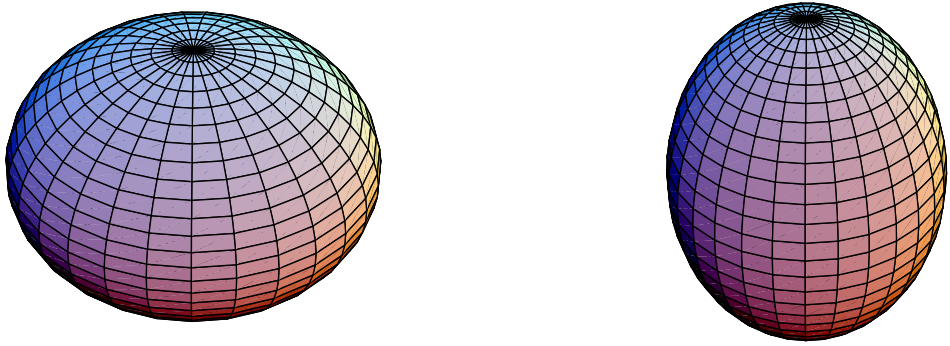 ,height=3in,width=6.5in }\caption{ A uniform
magnetic field or a cosmic string results in  an oblate spheroid
shape of the universe (left figure), while domain walls lead to
prolate spheroids (right figure) \cite{bere}. }
\end{figure}

  The energy-momentum tensor of the whole universe can be in general
given by
\begin{equation}
T^{\mu}_{\;\; \nu} = \mbox{diag} \,
(\rho,-p_{\|},-p_{\|},-p_{\bot}). \label{tensor}
\end{equation}
Then the Einstein equations read
\begin{eqnarray}
&&  \Bigl( \frac{\dot{a}}{a} \Bigr)^2 -\frac{2}{3}
\frac{\dot{a}}{a}\frac{e\dot{e}}{1-e^2}= \frac{8 \pi}{3} G \rho, \label{Ein a} \\
 && \Bigl(\frac{\dot{a}}{a} \Bigr)^2 +
\frac{2\ddot{a}}{a} -3\frac{\dot{a}}{a}\frac{e\dot{e}}{1-e^2}  -
\frac{e\ddot{e}}{1-e^2} -\frac{{\dot{e}}^2}{(1-e^2)^2}
= -8 \pi G p_{\|}, \label{Ein b}\\
&& \Bigl( \frac{\dot{a}}{a} \Bigr)^{2} + 2 \frac{\ddot{a}}{a} = -8
\pi G p_{\bot}, \label{Ein c}
\end{eqnarray}
where overdots denote derivatives with respect to the cosmic time.
>From Eqs. (\ref{Ein a}), (\ref{Ein b}) and (\ref{Ein c}), we get the
conservation equation for the energy-momentum tensor
\begin{equation}
\dot{\rho}+2\Bigl( \frac{\dot{a}}{a} \Bigr)
(\rho+p_{\|})+\Bigl(\frac{\dot{a}}{a} -2\frac{\dot{e}e}{1-e^2}
\Bigr)(\rho+p_{\bot})=0. \label{con eq}
\end{equation}
In principle, the total energy-momentum tensor, $T^{\mu}_{~\nu}$,
can be separated into two different parts: an anisotropic
contribution~$(T_{A})^{\mu}_{~\nu}=diag(\rho^{A},-p^{A}_{\|},
-p^{A}_{\|},-p^{A}_{\bot})$, which may include magnetic fields or
static aligned strings or static stacked walls, and an isotropic
contribution, $({T_{I}})^{\mu}_{~\nu}
=diag(\rho^{I},-p^{I},-p^{I},-p^{I})$, which includes symmetric
contributions from vacuum energy or  matter or radiation. As in the
isotropic universe, we may have a thermodynamic relation
\begin{equation}
\frac{Tdp^{I}}{dT}=\rho^{I}+p^{I},
\end{equation}
where $T$ is the temperature. Up to an additive constant, the
entropy in a volume $V$ is given by
\begin{equation}
S = (\rho^{I}+p^{I}) \frac{V}{T}.
\end{equation}
Using $V \propto \sqrt{1-e^{2}}a^{3}$, we obtain
\begin{equation}
\frac{\dot{S}}{S}
=(1-e^{2})^{-\frac{1}{2}}(-2e\dot{e})a^{3}+3a^{2}\dot{a}
(1-e^{2})^{\frac{1}{2}}+a^{3}(1-e^{2})^{\frac{1}{2}} \Bigl(
\frac{\dot{\rho}^{I}}{\rho^{I}+p^{I}} \Bigr). \label{ent ch}
\end{equation}
Assuming an adiabatic expansion of the universe, where the entropy
in a comoving volume is conserved, from Eq. (\ref{ent ch}) we obtain
a conservation equation
\begin{equation}
\dot{\rho}^{I}+3 \Bigl( \frac{\dot{a}}{a} \Bigr)(\rho^{I}+p^{I})
-\frac{2e\dot{e}}{1-e^{2}}(\rho^{I}+p^{I})=0. \label{is con eq}
\end{equation}
In the limiting case of $e=0$, Eq. (\ref{is con eq}) is exactly the
well-known conservation equation for the energy-momentum tensor in
the standard cosmology. For a matter with the equation of state
$p^{I}=\omega \rho^{I}$, by solving Eq. (\ref{is con eq}) we find
that the density evolves as
\begin{equation}
\rho^{I}\propto (1-e^{2})^{-(\omega+1)}a^{-3(\omega+1)}.
\end{equation}
Therefore, we have $\rho^{I}\propto (1-e^{2})^{-1}a^{-3}$ in the
matter-dominant era $(p^{I}=0)$ and $\rho^{I}\propto
(1-e^{2})^{-4/3}a^{-4}$ in the radiation-dominant era
$(p^{I}=\rho^{I}/3)$. Subtracting Eq. (\ref{is con eq}) for the
isotropic part from the conservation equation (\ref{con eq}) for the
total energy-momentum, we obtain a similar equation for the
anisotropic part
\begin{equation}
\dot{\rho}^{A}+2 \Bigl( \frac{\dot{a}}{a} \Bigr)
(\rho^{A}+p^{A}_{\|}) +
\Bigl(\frac{\dot{a}}{a}-\frac{2e\dot{e}}{1-e^{2}}
\Bigr)(\rho^{A}+p^{A}_{\bot})=0. \label{an con eq}
\end{equation}
Some exact solutions of the Einstein equations for kinds of plane
symmetric plus isotropic components are given in Ref. \cite{bere}.

We now give an example of a uniform magnetic field here, the
simplest case, whose physical interpretation in the brane-world
scenario will be discussed in Sec.~IV. Let us consider the
ellipsoidal universe in the matter-dominant era $(p^{I}=0)$, and
normalize the scale factors such that $a(t_{0})= c(t_{0})=1$ and
thus $e(t_{0})=0$ in the present time. The uniform magnetic field
has the energy-momentum tensor
${(T_{A})}^{\mu}_{~\nu}=\rho^{A}diag(1,-1,-1,1)$, where $\rho^{A} =
B^{2}/8\pi$ is the magnetic energy density. Here, it is assumed that
the magnetic field is frozen into the plasma due to the high
conductivity of the primordial plasma and the magnetic field evolves
as $B\propto a^{-2}$. Then, for a small eccentricity and thus
$(1-e^2)\sim 1$, Eqs. (\ref{Ein b}),~(\ref{Ein c}) and (\ref{an con
eq}) can be approximately written as
\begin{eqnarray}
~~~~~~~~~~&&\frac{d}{dt}(e\dot{e})+3He\dot{e}=16\pi G\rho^{A},\label{ecc eq}\\
~~~~~~~~~~&&\dot{H}^2+2\frac{\ddot{a}}{a}=8\pi G\rho^{A},\label{hub eq}\\
~~~~~~~~~~&&\dot{\rho}^{A}+4H\rho^{A}=0, \label{rho eq}
\end{eqnarray}where $H = \dot{a}/a$ is the Hubble constant
and $\Lambda$ is the cosmological constant. In the matter-dominant
era, we have $a(t)\sim (3H_{0}t/2)^{2/3}$ and then $H = 2/3t$. The
solution of Eq. (\ref{ecc eq}) is $e^2 = 8 \Omega^A_{(0)} ( 1 - 3
a^{-1} + 2 a^{-3/2})$, where $\Omega^A_{(0)} = \rho^A(t_0)/\rho_{\rm
cr}^{(0)}$, where $\rho_{\rm cr}^{(0)} = 3 H_0^2/8 \pi G$ is the
actual critical energy density. Since $a(t) < a(t_0) = 1$, the
dominant term of the eccentricity is $e^2 \sim 16 \Omega^A_{(0)}
a^{-3/2}$. The result of Ref. \cite{campanelli} shows that a small
eccentricity of order $e_{dec}\sim 10^{-2}$ at the decoupling epoch
generated by the uniform cosmic magnetic field with a strength
$B_{0}\sim 10^{-9}$G can explain the quadrupole problem without
affecting higher multipoles of the angular power spectrum of the
temperature anisotropy.

\section{Ellipsoidal universe in a bulk}

In the following, we discuss the ellipsoidal universe in the brane
world cosmology. The main purpose of this section is to derive the
corresponding equations governing the evolution of the ellipsoidal
universe for the brane metric.  In fact, recent results show that
brane world cosmological models have the capability to endow dark
energy with an excitingly new possibility ($\omega<-1$) without
suffering from the problems faced by phantom energy \cite{sah}. We
consider a 5D spacetime metric with an induced plane-symmetric
metric on the brane
\begin{equation}
ds^{2}=\tilde{g}_{AB}dx^{A}dx^{B}=g_{\mu\nu}dx^{\mu}dx^{\nu}-b^{2}dy^{2},
\end{equation}
where $y$ is the coordinate in the fifth dimension. Here and
hereafter, we focus our attention on the hypersurface defined by $y
= 0$, which we identify with the world volume of the brane that
forms our universe. The upper case Latin letters $A,B,...$ denote
5-dimensional indices, Greek letters $\mu,\nu,...$ denote indices
parallel to the brane world volume, 5 an index transverse to the
brane, and Latin letters $i,j,...$ denote space-like indices
parallel to the brane world volume. We can further take a metric of
the form
\begin{eqnarray}
ds^{2} &=& n^2(\tau,y)d
\tau^{2}-a^{2}(\tau,y)(dx_1^{2}+dx_2^{2})-\left(1-e^{2}(\tau,y)\right)
a^2(\tau,y)dx_3^{2}\nonumber\\&&-b^{2}(\tau,y)dy^{2}, \label{bra
met}
\end{eqnarray}
where the $x_1x_2$ plane is the plane of symmetry and $n (\tau,y)$
is a lapse function. The five-dimensional Einstein equations take
the usual form
\begin{equation}
\tilde{G}_{AB}\equiv
\tilde{R}_{AB}-\frac{1}{2}\tilde{R}\tilde{g}_{AB}=\kappa^{2}\tilde{T}_{AB},
\end{equation}
where $\tilde{R}_{AB}$ is the five-dimensional Ricci tensor,
$\tilde{R}=\tilde{g}^{AB}\tilde{R}_{AB}$ is the scalar curvature,
and the constant $\kappa$ is related to the five-dimensional
Newton's constant $G_{(5)}$ and the five-dimensional reduced Planck
mass $M_{(5)}$ by the relation, $\kappa^{2}=8\pi
G_{(5)}=M^{-3}_{(5)}$. The energy-momentum tensor can be decomposed
into two parts
\begin{equation}
\tilde{T}^{A}_{~B}=\check{T}^{A}_{~B}\mid_{bulk}+{T}^{A}_{~B}\mid_{brane},
\end{equation}
where $\check{T}^{A}_{~B}\mid_{bulk}$ is the energy-momentum tensor
of the bulk matter. The bulk tensor $\check{T}^{A}_{~B}\mid_{bulk}$
can be further decomposed into two parts: the isotropic
contribution,
$(\check{T_I})^{A}_{~B}\mid_{bulk}=\rho^I_{B}diag(-1,-1,-1,-1,-1)$,
from a cosmological constant, and the anisotropic part,
$(\check{T_\mathcal{A}})^{A}_{~B}\mid_{bulk} =
diag(\rho^\mathcal{A}_{B},
-P^{\mathcal{A}}_{B~\|},-P^\mathcal{A}_{B~\|},
-P^\mathcal{A}_{B~\bot},-P^\mathcal{A}_{B~T})$, where the bulk
energy density and pressures are independent of the coordinate $y$.
The second term ${T}^{A}_{~B}\mid_{brane}$ corresponds to the matter
on the brane($y=0$). The most general energy-momentum tensor
consistent with the planar symmetry takes the form
\begin{equation}
{T}^{A}_{~B}\mid_{brane}=\frac{\delta
(y)}{b}diag(\rho,-p_{\|},-p_{\|},-p_{\bot},0).
\end{equation}

Substituting Eq. (\ref{bra met}) into the Einstein equation, we get
the non-vanishing components of the Einstein tensor
$\tilde{G}^{A}_{~B}$:
\begin{eqnarray}
\tilde{G}^{0}_{I~0}&=&\frac{3}{b^{2}}\Bigl(-\frac{a'^{2}}{a^2}+\frac{a'b'}
{a b}+\frac{3ea'e'}{a (1-e^{2})}-\frac{a''}{a}\Bigr)
\nonumber\\&&+3\frac{\dot{a}}{an^{2}}\Bigl(\frac{\dot{a}}{a}
+\frac{\dot{b}}{b}\Bigr)
-\frac{e\dot{e}\dot{a}}{a n^{2}(1-e^2)}, \label{00}\\
\tilde{G}^{1}_{I~1}&=&G^{2}_{I~2}=\frac{1}{b^{2}}\Bigl(-\frac{a'^{2}}{a^{2}}
+\frac{2a'b'}{ab}+\frac{2ea'e'}{a(1-e^2)}
-2\frac{a'n'}{an}+\frac{b'n'}{bn}-\frac{2a''}{a}
-\frac{n''}{n}\Bigr)\nonumber\\
&&+\frac{1}{n^2}\Bigl(\frac{\dot{a}^{2}}{a^{2}}+\frac{2\dot{a}\dot{b}}{a
b}-\frac{2e\dot{a}\dot{e}}{a(1-e^2)}
-2\frac{\dot{a}\dot{n}}{an}-\frac{\dot{b}\dot{n}}{bn}
+\frac{2\ddot{a}}{a} +\frac{\ddot{b}}{b} \Bigr),\label{11}\\
\tilde{G}^{3}_{I~3}&=&
\frac{1}{b^{2}}\Bigl(-\frac{a'^{2}}{a^{2}}+\frac{2a'b'}{a
b}-2\frac{a'n'}{an}+\frac{b'n'}{nb} -\frac{2a''}{a
}-\frac{n''}{n}\Bigr)\nonumber\\&&
+\frac{1}{n^{2}}\Bigl(\frac{\dot{a}^2}{a^2}+\frac{2\dot{a}\dot{b}}{a
b}-\frac{2\dot{a}\dot{n}}{an}-\frac{\dot{b}\dot{n}}{bn}
+\frac{2\ddot{a}}{a}+\frac{\ddot{b}}{b}\Bigr),\label{33}\\
\tilde{G}^{0}_{I~5}&=&
\frac{1}{n^{2}}\Bigl(\frac{3n'\dot{a}}{an}+\frac{3a'\dot{b}}{a
b}-\frac{e a'\dot{e}}{a(1-e^2)}+\frac{3e e'\dot{a}}{a(1-e^2)}
-\frac{3\dot{a}'}{a}\Bigr),\label{05}\\
\tilde{G}^{5}_{I~0}&=&\frac{1}{b^{2}}\Bigl(-\frac{3n'\dot{a}}{an}-\frac{3a'\dot{b}}{a
b} +\frac{e a'\dot{e}}{a(1-e^2)}-\frac{3e e'\dot{a}}{a(1-e^2)}
+\frac{3\dot{a}'}{a}\Bigr),\label{50}\\
\tilde{G}^{5}_{I~5}&=&\frac{1}{n^{2}}\Bigl(\frac{3\dot{a}^2}{a^2}-\frac{3
e\dot{e}\dot{a}}{a(1-e^2)}-\frac{3\dot{a}\dot{n}}{an}
+\frac{3\ddot{a}}{a}\Bigr)
\nonumber\\&&+\frac{1}{b^{2}}\Bigl(-\frac{3a'^{2}}{a^{2}}+\frac{e a'
e'}{a (1-e^2)}-\frac{3a'n'}{an}\Bigr). \label{55}
\end{eqnarray}
In the above expressions, primes stand for derivatives with respect
to $y$, and overdots for derivatives with respect to the cosmic time
$\tau$. The reason why we can separate the Einstein tensors into
isotropic parts and anisotropic parts is that the corresponding
parts of the energy-momentum tensor obey their own conservation
equations, Eqs. (\ref{is con eq}) and (\ref{an con eq}). We assume
that there is no flow of matter along the fifth dimension, i.~e.
$(\check{T}_I)_{05}=0$, which in turn implies that
$\tilde{G}^{0}_{I~5~}=\tilde{G}^{5}_{I~0}=0$. Then the components
$(0,0)$ and $(5,5)$ of Einstein equations (\ref{00}) and (\ref{55})
in the bulk can be written in the simple form
\begin{eqnarray}
F'=\frac{2a'a^3}{3}(1-e^2)\kappa^{2}\check{T}^{0}_{I~0},~~~
    \dot{F}=\frac{2\dot{a}a^3}{3}(1-e^2)\kappa^{2}\check{T}^{5}_{I~5},
\end{eqnarray}
where $F$ is a function of $\tau$ and $y$ defined by
\begin{equation}
F(\tau,y) = (1-e^2) \Bigl[ \Bigl(\frac{\dot{a}a}{n} \Bigr)^2 -
\Bigl( \frac{a'a}{b} \Bigr)^2 \Bigr]. \label{f}
\end{equation}
Integrating Eq. (\ref{f}), we obtain the relation
\begin{equation}
\Bigl( \frac{\dot{a}}{an} \Bigr)^2 = \Bigl( \frac{a'}{ab} \Bigr)^2 +
                          \frac{\kappa^2}{6}\check{T}^{0}_{I~0}-\frac{\int
a^4d(1-e^2)}{6a^4(1-e^2)}\check{T}^{0}_{I~0}
+\frac{\mathcal{C}_{1}}{a^{4}(1-e^2)},\label{ae rel}
\end{equation}
where $\mathcal{C}_{1}$ is a constant of integration. The above
equation shows that the scale factor $a(\tau, y)$ and the
eccentricity are coupled to each other. If $e=0$, then Eq. (\ref{ae
rel}) reduces exactly to the result of Ref. \cite{bin}.

We now consider the anisotropic contribution of the total
energy-momentum tensor. The corresponding anisotropic part of the
Einstein tensors can be written as
\begin{eqnarray}
\tilde{G}^{0}_{\mathcal{A}~0}&=&\frac{1}{b^{2}(1-e^{2})}\Bigl({ee''}
+\frac{e'^2}{(1-e^{2})}-\frac{eb'e'}{b} +\frac{ea'e'}{a}\Bigr)
\nonumber\\&&-\frac{1}{n^{2}(1-e^2)}
\Bigl(\frac{e\dot{a}\dot{e}}{a}+\frac{e\dot{b}\dot{e}}{b}\Bigr),\label{a00}\\
\tilde{G}^{1}_{\mathcal{A}~1}&=&
\tilde{G}^{2}_{\mathcal{A}~2}=\frac{1}{b^{2}(1-e^{2})}\Bigl(\frac{ea'e'}{a
}-\frac{e b' e'}{ b}+\frac{e'^2}{(1-e^{2})}+ee'n'+{e
e''}\Bigr)\nonumber\\&&-\frac{1}{n^2(1-e^2)}\Bigl(\frac{e\dot{a}\dot{e}}{b}
+\frac{e\dot{b}\dot{e}}{b}+\frac{\dot{e}^2}{(1-e^2)}
-\frac{e\dot{e}\dot{n}}{n}+\ddot{e}e\Biggr),\label{a11}\\
\tilde{G}^{3}_{\mathcal{A}~3}&=& 0,\label{a33}\\
\tilde{G}^{0}_{\mathcal{A}~5}&=&\frac{1}{n^2(1-e^2)}\Bigl(-\frac{e
e'\dot{a}}{a}-\frac{e e' \dot{b}}{b}+\frac{e a'
\dot{a}}{a}+\frac{e'\dot{e}}{(1-e^2)}-\frac{en'\dot{e}}{n}
+e\dot{e}'\Bigr),\label{a05}\\
\tilde{G}^{5}_{\mathcal{A}~0}&=&\frac{1}{b^{2}(1-e^{2})}\Bigl(\frac{e
e'\dot{a}}{a }+\frac{e e' \dot{b}}{b}-\frac{e a' \dot{a}}{a }-
\frac{e'\dot{e}}{(1-e^2)}+\frac{en'\dot{e}}{n}-
{e\dot{e}'}\Bigr),\label{a50}\\
\tilde{G}^{5}_{\mathcal{A}~5}&=& \frac{1}{1-e^2}\Bigl(\frac{ea'e'}{a
b^2}+\frac{ee'n'}{b^2}-\frac{e\dot{a}\dot{e}}{an^2}
-\frac{\dot{e}^2}{n^2}+\frac{e\dot{e}\dot{n}}{n^3}
-\frac{e\ddot{e}}{n^2}\Bigr). \label{a55}
\end{eqnarray}
Here we also assume that $\check{T}^{0}_{\mathcal{A}~5}=0$, that is
to say no flow of anisotropic matter (magnetic fields or cosmic
strings or domain walls) along the fifth dimension and so
$G^{0}_{\mathcal{A}~5}=G^{5}_{\mathcal{A}~0}=0$. Eqs. (\ref{a00})
and (\ref{a55}) can be written as
\begin{eqnarray}
K' \simeq 2e
e'(1-e^2)a^2\kappa^{2}\check{T}^{0}_{\mathcal{A}~0},~~~\dot{K}\simeq
2 e \dot{e}(1-e^2)a^2\kappa^{2}\check{T}^{5}_{\mathcal{A}~5},
\end{eqnarray}
where $K$ is defined by
\begin{equation}
K = \Bigl( \frac{a e e'}{b} \Bigr)^2 - \Bigl( \frac{a e \dot{e}}{n}
\Bigr)^2.
\end{equation}
Integrating Eq.  (38) we obtain
\begin{equation}
\Bigl( \frac{e \dot{e}}{(1-e^2)n} \Bigr)^2 = \Bigl(\frac{e
e'}{(1-e^2)b} \Bigr)^2 +
\frac{1}{2}\kappa^{2}\check{T}^{0}_{\mathcal{A}~0}+\frac{\int
(1-e^2)da^2}{a^2(1-e^2)}\check{T}^{0}_{\mathcal{A}~0}
+\frac{\mathcal{C}_{2}}{a^2(1-e^2)^2},
\end{equation}
where $\mathcal{C}_{2}$ is another integration constant.

We now take the brane into consideration by using the Israel's
junction condition \cite{isr}. When the coordinate system (18) is
chosen, the extrinsic curvature tensor of a given hypersurface (for
example, the $y=0$ surface) is defined by $K_{\mu\nu} =
\partial_{y}g_{\mu\nu}/2$. The Israel's junction condition at $y=0$ can be
written as \cite{deffe,bridg}
\begin{equation}
\{K_{\mu\nu}-Kg_{\mu\nu}\}_{y}=-\frac{\kappa^2}{2}T_{\mu\nu},
\end{equation}
where $K$ is defined by $K\equiv K_{\mu\nu}g^{\mu\nu}$, $T_{\mu\nu}$
is the energy-momentum tensor of the brane. We have assumed a brane
with $Z_{2}$ symmetry and the index $_{y}$ means accordingly that
the value of the component $K_{\mu\nu}$ is taken over one side of
the brane, namely at $y=0^{+}$.  The junction condition on the
planar symmetric metric background are given by
\begin{eqnarray}
\frac{[a']}{a_0b_0} &=&\frac{\kappa^2}{3}\rho
+\frac{\kappa^2}{3}(p_{\|}-p_{\bot}),\\
\frac{e_{0}[e']}{b_0(1-e_0^2)}&=&\kappa^2(p_{\|}-p_{\bot}),\\
\frac{[n']}{n_0b_0}&=&-\frac{\kappa^2}{3}(2\rho+2p_{\|}+p_{\bot})
\end{eqnarray}
where the subscript $0$ for $a$, $b$, and $n$ means that they are
taken at $y=0$, and $[Q]=Q(0^{+})-Q(0^{-})$ denotes the jump of the
function $Q$ across $y=0$. From Eqs.  (42)-(44), we can see that
when $p_{\|}=p_{\bot}$ the boundary conditions reduce to the
spherical cases discussed in Ref. \cite{bin}. Assuming the symmetry
$y \leftrightarrow -y$ for simplicity, the junction conditions
(42)-(44) can be used to compute $a'$ and $e'$ on the two sides of
the brane, and by continuity when $y \rightarrow  - y$, Eq. (31) and
Eq. (40) are reexpressed as (after setting $n_0=1$)
\begin{eqnarray}
&&\Bigl( \frac{\dot{a}}{a_0} \Bigr)^2=
\frac{\kappa^2\rho_B^I}{6}+\frac{\kappa^4\rho^2}{36}
+\frac{\kappa^4}{36}(p_{\|}-p_{\bot})^2-\frac{\int
a^4d(1-e^2)}{6a_0^4(1-e_0^2)}\rho_B^I+\frac{\mathcal{C}_{1}}{a_0^{4}(1-e_0^2)}~~\\
&&\Bigl( \frac{{e_0}{\dot{e}}_0}{1-{e_0}^2} \Bigr)^2=
\frac{\kappa^4}{4}(p_{\|}-p_{\bot})^2
+\frac{\kappa^2}{2}\rho_B^{\mathcal{A}}+\frac{\int
(1-e^2)da^2}{a_0^{2}(1-e_0^2)^2}\rho_B^{\mathcal{A}}+\frac{\mathcal{C}_{2}}{a_0^{2}(1-e_0^2)^2}.
\end{eqnarray}
The above two equations are the main results of our work. From Eq.
(45), we can see that when $p_{\|}=p_{\bot}$, the scale factor is
exactly what was found in Ref. \cite{bin}. If the third term in Eq.
(45) is small compared with the other terms, it play a role of
perturbation and the Friedman equation does not deviate much from
that of Ref. \cite{bin}. Eq. (46) describes the evolution of the
eccentricity on the brane world, which shows that the evolution of
the eccentricity depends on both the brane anisotropic pressures and
the bulk anisotropic energy density. We should notice that the
energy conservation on the brane, which was obtained in Sec. II,
still works here
\begin{equation}
\dot{\rho}+2 \Bigl( \frac{\dot{a}}{a} \Bigr) (\rho+p_{\|}) +
\Bigl(\frac{\dot{a}}{a}-2\frac{\dot{e}e}{1-e^2}
\Bigr)(\rho+p_{\bot})=0.
\end{equation}

Since the scale factor and the eccentricity are coupled to each
other, we do not expect to solve Eqs. (45) and (46) exactly. But for
a small eccentricity, i.e. $\rho\sim
(1-e^2)^{-(\omega+1)}a^{-3(\omega+1)}\approx a^{-3(\omega+1)}$ and
when the third term in Eq. (45) can be neglected, the time evolution
of the scale factor can be given by \cite{bin}
\begin{equation}
a_{0}(t)=a_{*}(\kappa^2\rho_{*})^{1/q}\Bigl(
\frac{q^2}{72}\kappa^2\rho_{\Lambda}t^2+\frac{q}{6}t \Bigr)^{1/q},
\end{equation}
where the energy density on the brane has been decomposed into three
parts
 $\rho=\rho^I+\rho^{\mathcal{A}}+\rho_{\Lambda}$,
 $q=3(\omega+1)$, $a_{*}$ and  $\rho_{*}$ are constants,
 and $\rho_{\Lambda}$ is a constant that represents
an intrinsic tension of the brane. Here the Randall-Sundrum
relation, $\kappa^2 \rho_B^I/ 6 + \kappa^2 \rho_{\Lambda}^2/36 = 0$
has been used \cite{randall}. Eq. (48) indicates that at a very
early universe, the cosmology is characterized by $a(t)\sim
t^{1/q}$, while at a late time it is described by the standard
cosmology, $a \sim t^{2/q}$. We are going to solve Eq. (46), by
considering only the first term and neglecting others. We still
assume the anisotropy of the our universe is contributed by a
uniform magnetic field with the energy-momentum tensor
${(T_{\mathcal{A}})}^{\mu}_{~\nu}=\rho^{{\mathcal{A}}}diag(1,-1,-1,1)$,
and thus from Eq. (47) we find that $\rho^{{\mathcal{A}}}\propto
a^{-4}$. Substituting $a \sim t^{1/q}$ , $H = 1/qt$, and
$\rho^{{\mathcal{A}}}= \rho_0^{{\mathcal{A}}}a^{-4}$ back to Eq.
(46), we obtain
\begin{equation}
e^2 = \frac{2\kappa^2
\rho_0^{{\mathcal{A}}}t^{1-\frac{4}{q}}}{q-4}+\mathcal{C}_3,
\end{equation}
where $\mathcal{C}_3$ is an integration constant. Similarly, at a
late time of the universe, $a \sim t^{2/q}$ , $H = 2/q t$, and
$\rho^{A}= \rho_0^{{\mathcal{A}}}a^{-4}$, the eccentricity is
expressed as
\begin{equation}
e^2 = \frac{2\kappa^2
\rho_0^{{\mathcal{A}}}t^{1-\frac{8}{q}}}{q-8}+\mathcal{C}_4,
\end{equation}
where $\mathcal{C}_4$ is also another integration constant. In the
matter-dominant era $(q=3)$, we find that the late-time evolution of
eccentricity in the bulk approximately is $e^2\sim a^{- 5/2}$, which
decays a little faster than that in 4D universe, while $e^2\sim
a^{-1}$ for a very early universe $a \sim
t^{1/q}$ with $q=3$.\\

\section{Ellipsoidal Universe in the brane-induced gravity model}

In this section, we explore a scenario of the ellipsoidal universe
based on the Dvali-Gabadadze-Porrati (DGP) model of brane-induced
gravity \cite{dav}. In this model, the 3-brane is embedded in a
spacetime with an infinite-size extra dimension. The usual
gravitational laws is obtained by adding to the action of the brane
an Einstein-Hilbert term computed with the intrinsic curvature on
the brane. Particularly, one recovers a standard four-dimensional
(4D) Newtonian potential for small distances, whereas gravity is in
a 5D regime for large distances. The cosmology of this model in the
case of a 5D bulk was studied by Daffayet, Dvali and Gabadadze. It
is shown there that if the cosmological model contains a scalar
curvature term in the action for the brane, besides the brane and
bulk cosmological constraints, the presence of the scalar curvature
term in the brane action can lead to a late-time acceleration of the
universe even in the absence of any material form of dark energy
\cite{deff,def}.

Following the framework of Refs. \cite{deff,def}, we consider a
3-brane embedded in a 5D spacetime with an intrinsic curvature term
induced on the brane and the action of the form
\begin{equation}
S_{(5)}=-\frac{1}{2\kappa^{2}}\int
d^{5}x\sqrt{-\tilde{g}}\tilde{R}+\int d^{5}x
\mathcal{L}_{m}-\frac{1}{2\mu^{2}}\int d^{4}x \sqrt{-g}R
\end{equation}
The first term in Eq. (51) corresponds to the Einstein-Hilbert
action in five dimensions for a 5D metric $\tilde{g}_{AB}$(bulk
metric). Similarly, the last term in (51) is the Einstein-Hilbert
action for the induced metric $g_{cd}$ on the brane, $R$ being its
scalar curvature. The induced metric $g_{cd}$ is defined as usual
from the bulk metric,
$g_{cd}=\partial_{c}X^{A}\partial_{d}X^{B}\tilde{g}_{AB}$, where
$X^{A}(x^{c})$ represents the coordinates of an event on the brane
labeled by $x^{c}$. The second term in (51) denotes the matter
content.

The five-dimensional Einstein equations in the brane-induced gravity
is given by
\begin{equation}
\tilde{G}_{AB}\equiv
\tilde{R}_{AB}-\frac{1}{2}\tilde{R}\tilde{g}_{AB}=\kappa^{2}\tilde{S}_{AB},
\end{equation}
where the tensor $\tilde{S}$ is the sum of the energy-momentum
tensor $\tilde{T}$ of matter and the contribution coming from the
scalar curvature of the brane. We denote the latter contribution
$\tilde{U}$, so
\begin{equation}
\tilde{S}^{A}_{~B}=\tilde{T}^{A}_{~B}+\tilde{U}^{A}_{~B}.
\end{equation}
The junction condition in the brane-induced gravity model is
replaced by $ \{K_{\mu\nu} - Kg_{\mu\nu}\}_{q} = - (\kappa^2 / 2)
S_{\mu\nu}, $ with $S_{\mu\nu} = T_{\mu\nu} - (1 / \mu^2)
U_{\mu\nu}$, where $\mu^2$ is the four-dimensional Newton's
constant, $\mu^2=8\pi G_{(4)}=M_{(4)}^{-2}$. The Israel boundary
condition is then written as
\begin{eqnarray}
\frac{[a']}{a_0b_0}&=& \frac{\kappa^2}{3}\rho
+\frac{\kappa^2}{3}(p_{\|}+p_{\bot}) -\frac{\kappa^2}{3\mu^2n^2_0}
\nonumber\\&&\Bigl(3\frac{\dot{a}^2}{a^2_0}
 +\frac{e_0\dot{a
}\dot{e}}{a_0
(1-e^2_0)}+\frac{e_0\ddot{e}}{1-e^2_0}-\frac{e_0\dot{e}\dot{n}}{n_0(1-e^2_0)}
+\frac{\dot{e}^2}{(1-e^2_0)^2} \Bigr), \label{jun a}
\\
\frac{e_0[e']}{b_0(1-e^2_0)}&=&\kappa^2(p_{\|}-p_{\bot})\nonumber\\&-&
\frac{\kappa^2}{\mu^2n^2_0} \Bigl(\frac{3e_0\dot{a }\dot{e}}{a_0
(1-e^2_0)}+\frac{e_0\ddot{e}}{1-e^2_0}
-\frac{e_0\dot{e}\dot{n}}{n_0(1-e^2_0)}+\frac{\dot{e}^2}{(1-e^2_0)^2}
\Bigr),
\label{jun b} \\
\frac{[n']}{n_0b_0}&=&-\frac{\kappa^2}{3}(2\rho+2p_{\|}+p_{\bot})
+\frac{\kappa^2}{\mu^2n^2_0} \Bigl(\frac{\dot{a}^2}{a^2_0}
+2\frac{\dot{a}\dot{n}}{a_0n_0} -\frac{2\ddot{a}}{a_0} \Bigr)
+\frac{2\kappa^2}{3\mu^2n^2_0}\nonumber\\&& \Bigl(\frac{e_0\dot{a
}\dot{e}}{a_0 (1-e^2_0)} +\frac{e_0\ddot{e}}{1-e^2_0}
-\frac{e_0\dot{e}\dot{n}}{n_0(1-e^2_0)}
+\frac{\dot{e}^2}{(1-e^2_0)^2}\Bigr)~.~ \label{jun c}
\end{eqnarray}
Substituting Eqs. (31), (40), (54) and (55) into the (0,0)-component
of the Israel's junction condition and assuming $\rho_B^I$,
$\rho_B^{\mathcal{A}}$, $\mathcal{C}_{1}$ and $\mathcal{C}_{2}$ to
be approximately zero, we have
\begin{equation}
\Bigl( H-\epsilon\frac{\mu^2}{\kappa^2}
\Bigr)^{2}-\frac{2}{3}\frac{e\dot{e}}{1-e^2}
\Bigl(H-\epsilon\frac{\mu^2}{\kappa^2} \Bigr) \simeq
\frac{\mu^2}{3}\rho+\epsilon^2\frac{\mu^4}{\kappa^4},
\end{equation}
where $H = \dot{a}/a_0$, $\epsilon$ is the sign of $[a']$ such that
$\epsilon=\pm 1$ and the lapse function is set to be $n_0=1$. One
should note that when $\mu^2 / \kappa^2 < 2 \mu^2 / \kappa^2 \ll H$,
or in terms of the Hubble radius $H^{-1}$, when $H^{-1}\ll M_{(4)}^2
/ 2M_{(5)}^3$, the 4D ellipsoidal universe is recovered, i.e. one
can return to Eq. (5).

We notice that from Eqs. (40) and (55) with vanishing anisotropic
bulk energy-momentum tensor $ \rho_B^{\mathcal{A}}$ and vanishing
$\mathcal{C}_{2}$, it follows that
\begin{equation}
\Bigl(\frac{e\ddot{e}}{1-e^2}+\frac{\dot{e}^2}{(1-e^2)^2}\Bigr)+
{\frac{(e\dot{e})}{(1-e^2)}}
\Bigl(3\frac{\dot{a}}{a}+2\epsilon\frac{\mu^2}{\kappa^{2}}\Bigr)={\mu^2}
(p_{\|}-p_{\bot}).
\end{equation}
When $2\epsilon \mu^2/ \kappa^2$ in the second term can be
neglected, then Eq. (58)  exactly equals to Eq. (\ref{Ein c})
subtracted by Eq. (\ref{Ein b}). In the small eccentricity
approximation $1-e^2\sim 1$, we solve Eq. (58) in the
matter-dominant era  with $a\sim t^{2/3}$ and $H = 2/3t$. We
normalize the scale factor such that $a(t_{0})=1$ and $e(t_{0})=0$
in the present time and consider the anisotropy on the brane
contributed by a uniform magnetic field with the energy-momentum
tensor
${(T_{{\mathcal{A}}})}^{\mu}_{~\nu}=\rho^{{\mathcal{A}}}diag(1,-1,-1,1)$.
The solution of Eq. (58) can be approximately written as
\begin{eqnarray}
e^2(t)& \sim &\int_{t}^{t_0}
4\pi\rho_0^{A}e^{-2\epsilon\frac{\mu^2}{\kappa^2}t'}t'^{-2}dt'\nonumber\\
&=&\int_{t}^{t_0} 4\pi\rho_0^{A}
\Bigl(1-2\epsilon\frac{\mu^2}{\kappa^2}t'
+\frac{4\frac{\mu^4}{\kappa^4}t'^2}{2!}+ \cdots \Bigr)t'^{-2}dt'.
\end{eqnarray}
Eq. (59) indicates that $\epsilon$ should be $1$, otherwise the
eccentricity in the brane does not converge when $t\rightarrow
\infty$. Fortunately, Ref. \cite{deff} shows that the brane
cosmology with $\epsilon=1$ can produce a late-time accelerated
expansion. When $2 \mu^2 / \kappa^2 \ll H$ and then $
e^{-2\epsilon\frac{\mu^2}{\kappa^2}t}\simeq
e^{-2\epsilon\frac{\mu^2}{\kappa^2}H^{-1}}\sim 1$, thus  $e^2(t)\sim
t^{-1}\sim a^{- 3/2}$, which is exactly the dominant term of the 4D
case. Integrating Eq. (59), the eccentricity on the brane can be
written as
\begin{equation}
\label{e2} e^2(t)\simeq \frac{1}{3}\Omega^{\mathcal{A}}_{(0)}
\Bigl(a^{-3/2}+2 \frac{\ln a}{r_{c}} - \frac{a^{3/2}}{r_{c}^2}
\Bigr),
\end{equation}
where $\Omega^{\mathcal{A}}_{(0)} =
\rho^{\mathcal{A}}(t_0)/\rho_{\rm cr}^{(0)}$, $\rho_{\rm cr}^{(0)} =
3 H_0^2/8 \pi G$ and $r_{c} = \kappa^2 / 2\mu^2 = M_{(4)}^2 /
2M_{(5)}^3$ is the length scale for the crossover between 4D gravity
and the 5D gravity regimes \cite{dav}. Differently from that of the
4D case, the brane-induced gravity contributes some higher order
corrections to the eccentricity duo to the presence of an extra
dimension.

\section{Discussion and Conclusion}

In the brane cosmology we have found the Friedmann-like equation for
the ellipsoidal universe in the bulk, which depends on the geometry
and matter (both isotropic and anisotropic) content  of the brane.
We also have found that the evolution equation of the eccentricity
in the bulk depends only on the anisotropic pressures inside the
brane and the anisotropic energy density in the bulk, except a
constant parameter. The evolution equation of the eccentricity is
coupled to that of the scale factor. As a model calculation, we have
shown that if only a uniform magnetic field inside the brane
contributes to the anisotropy of our universe but the anisotropic
energy density from the bulk and other terms are neglected, the
evolution of the eccentricity decays faster than that in a 4D
universe. To compare with the 4D ellipsoidal universe, we have
considered the ellipsoidal universe with a 3-brane embedded in a 5D
spacetime with an intrinsic curvature term included in the brane
action. The results show that the usual ellipsoidal cosmology is
recovered for Hubble radii smaller than the crossover scale given by
$r_c={M_{(4)}^2}/{2M_{(5)}^3}$ between 4D and 5D gravity.

We now briefly discuss the relation between eccentricity and cosmic
microwave background quadrupole problem. For a small eccentricity of
the universe in the brane, the metric tensor may be written in a
perturbation form
 \begin{equation}
 ds^2=dt^2-a^2(t,y)(\delta_{ij}-h_{ij})dx^{i}dx^{j}-b^2(t,y)dy^2,
 \end{equation}
 where $h_{ij}$ represents a metric perturbation,
 $h_{ij} = e^2 \delta_{i3} \delta_{j3}$.
 Suppose a photon emitted at $t=t_{dec}$ from
 the last scattering surface travels along a null geodesic
 and reaches an observer at $t=t_0$. Let $n^{i}$ be a unit vector
 along the null ray from the observer to the surface.
 With respect to the observer at the origin, the photon ray
 is  $x^{i} (t) =\eta(t)n^{i}$, where
 $\eta(t)=\int_{t}^{t_0} dt'/ a(t')$.
 The geodesic deviation between an observer at $x^i$ and another
 observer at $x^i+\delta x^i$ along the ray at any time $t$
 is given by $\delta l = (-g_{ij}\delta x^i\delta x^j) \propto
 a(1 - h_{ij}n^{i}n^{j}/2)$ to first order in $h_{ij}$.
>From the proper velocity $v = d \delta l/dt$ of one observer with
respect to another and the redshift of the photon's physical
frequency $\delta \omega_p / \omega_p = -v$, we find the redshift of
the comoving frequency
 \begin{equation}
\frac{\delta \omega_c}{ \omega_c} = \frac{\delta (\omega_p a)}{
\omega_p a}=\frac{1}{2}\frac{\partial h_{ij}}{\partial t}n^{i}n^{j}
\delta t. \label{red sh}
 \end{equation}
As the temperature of radiation with respect to the comoving
observer is proportional to the frequency $(\delta T \propto
\omega_c )$, we may find the temperature variation from Eq.
(\ref{red sh}) as
\begin{equation}
\frac{\delta T}{ T_0}=\frac{1}{2}\int_{t_{dec}}^{t_0}\frac{\partial
h_{ij}}{\partial t}n^{i}n^{j} d t=\frac{1}{2}e^2_{\rm dec}n^2_3,
 \end{equation}
 where $\label{nA} n_{\rm 3}(\theta,\phi) = \cos \theta \cos \vartheta -
\sin \theta \sin \vartheta \cos(\phi - \varphi)$  and
$\vartheta,\varphi$ are the angles between spherical galactic
coordinates and the c-axis and a-axis respectively
\cite{campanelli}.

On the other hand, the relative temperature anisotropy $\delta
 T(\theta,\phi)/T_0$ leads to the power spectrum
\begin{equation}
\frac{\delta T_l}{\langle T \rangle} = \sqrt{ \frac{1}{2 \pi} \,
\frac{l(l+1)}{2l+1} \sum_m |a_{lm}|^2}, \label{power sp}
\end{equation}
where $a_{lm}$ is the coefficient of spherical harmonics
$Y_{lm}(\theta,\phi)$ and $\langle T \rangle = 2.726 \pm 0.010 K$ is
the actual average temperature of the CMB radiation. The power
spectrum  (\ref{power sp}) fully describes all the CMB anisotropy
and $l=2$ refers to the quadrupole anisotropy. \textsl{Recent WMAP
data hints a violation of statistical isotropy on its largest scales
and a missing power at scales greater than $60^{\circ}$
\cite{Spergel}.} Particularly, the observed quadrupole anisotropy
\begin{equation}
(\delta T_2)^2_{obs}\simeq 236\mu K^2,
\end{equation}
deviates by order of magnitude from the prediction of standard
inflation
\begin{equation}
(\delta T_2)^2_{I}\simeq 1252\mu K^2.
\end{equation}
This anomaly is called the quadrupole problem. Since we have assumed
that the large-scale spatial geometry of our universe on the brane
is plane symmetric with a small eccentricity, the observed CMB
anisotropy map is a linear superposition of two independent
contributions $\delta T= \delta T_{A}+\delta T_{I} $, where $\delta
T_{A}$ represents the temperature fluctuations due to the
anisotropic spacetime background, while $\delta T_{I}$ is the
standard isotropic fluctuation caused by the inflation-led
gravitational potential at the last scattering surface. Similarly,
we may write $a_{lm}=a^{A}_{lm}+a^{I}_{lm}$, where $a^{A}_{lm} =
(1/2) \int_{0}^{2\pi}\int_{0}^{\pi} e^2_{\rm dec}n^2_3 Y^{*}_{lm}
sin \theta d\theta d\phi$. Finally we note that while the quadrupole
anisotropy contributed by the uniform magnetic field is
$\mathcal{S}_{A} = \delta T_2/ \langle T \rangle = (2/5\sqrt{3})
e^2_{\rm dec}$ with an eccentricity of order $10^{-2}$, the
predicted quadrupole anisotropy can be in a range of $46.2 \mu
\mbox{K}^2 \leq\left(\delta T_2\right)^{2}\leq 1001.6 \, \mu
\mbox{K}^2$ (for detailed calculations, see Ref. \cite{campanelli}).
Thus the data is in agreement with observations.

The equation that governs the evolution of eccentricity in a
five-dimensional bulk is found in Sec. III, which is shown to depend
on the anisotropic pressures inside the brane and the anisotropic
energy density in the bulk. If  anisotropy of the our universe is
contributed by a uniform magnetic field inside the brane, and
neglecting the anisotropic energy density from the bulk and other
terms, the evolution of the eccentricity decays faster than that in
a 4D ellipsoidal universe. Since the evolution of eccentricity here
is described by a 5-dimensional Newton's constant and is
inconvenient to be compared with observational data, we have come to
focus on the case of ellipsoidal universe in the brane-induced
gravity in Sec. IV. The perturbational metric takes the form of Eq.
(61) where we have assumed the fifth dimension to be stabilized
under the perturbation. This is because the origin of eccentricity
is contributed by the magnetic field which according to
Ho$\rm{\check{r}}$ava-Witten picture is confined to a 3-brane
\cite{ha}. From Eq. (60), we can determine the value of $e^2_{\rm
dec}$. At the decoupling, $t=t_{\rm dec}$, we have $e^2_{\rm
dec}\simeq \frac{1}{3}\Omega^{\mathcal{A}}_{(0)}(z^{3/2}_{\rm
dec}-2\frac{\rm{ln} z_{\rm dec}}{r_{c}}-\frac{z^{-3/2}_{\rm
dec}}{r^2_{c}})$, where $z_{\rm dec}\simeq 1088$ is the redshift at
decoupling \cite{Spe}. For large $r_{c}$ (for example,
$r_{c}=1.21^{+0.09}_{-0.09}H^{-1}_0$~\cite{deff}), we have $e^2_{\rm
dec}\simeq \frac{1}{3}\Omega^{\mathcal{A}}_{(0)}z^{3/2}_{\rm dec}$.
Finally, we get $e_{\rm dec}\simeq
\frac{1}{4\sqrt{3}}10^{-2}h^{-1}\frac{B_0}{10^{-8}G}$, where
$B_0=B(t_0)$ is defined by $\rho^{{\mathcal{A}}}_0=B^2_0/8\pi$, and
$h\simeq 0.72$. Thus, for $B_{0}\simeq (4-5)\times 10^{-9}$G, we
have $e_{\rm dec}\simeq(0.1-0.2)\times 10^{-2}$. Therefore, in the
brane-induced gravity model, the evolution of eccentricity in the
brane can also be used to explain the quadrupole problem of the
cosmic microwave background.

In summary, we have considered a version of ellipsoidal universe in
the brane world scenario. Here, we emphasize that the ellipsoidal
universe discussed above is undergoing a homogenous but anisotropic
expansion. That is to say, before the onset of inflation, a typical
region of the universe is homogeneous and isotropic but in some
regions an asymmetric expansion driven by magnetic fields would have
stretched out the regions and left with an imprint through the end
of inflation \cite{bere}. The ellipsoidal universe has been used to
explain the suppression of quadrupole moment, but not all the low
multipoles since in the ellipsoidal coordinate all the coefficients
$a^{A}_{lm}$ with $l> 2$ vanish. This can be easily found by
calculating the formula $a^{A}_{lm} = (1/2)
\int_{0}^{2\pi}\int_{0}^{\pi} e^2_{\rm dec}n^2_3 Y^{*}_{lm} sin
\theta d\theta d\phi$. In the above discussions, we have also
investigated the ellipsoidal universe in the DGP model and used a
large $r_{c}$ in obtaining the value of eccentricity. We notice that
up to now the comparison between data and the DGP model based upon
the expansion history of the universe is still contradictory.
 Recent data of Supernova and CMB suggested that the
self-accelerating branch of the DGP model is somehow disfavored
\cite{ish,amar,sawi,gor,confliction,as,song}, while the most recent
analysis of the new 'gold' data set of supernovae \cite{ries} and
the CMB shift parameter suggested that a flat universe is completely
consistent with the DGP model \cite{bar,alam}. However, as the
density perturbations of the DGP model may differ from those of
$\Lambda$CDM model, it is still unclear at present whether the DGP
model is marginally or significantly disfavored or not
\cite{sara,linder,ma}. A full examination of this issue is beyond
the scope of this paper, but from Eq. (\ref{e2}) we can see that a
large enough $r_{c}$ contributes little to the value of the
eccentricity.

\ack

X.~H.~G. would like to thank Prof. C.~Deffayet for his helpful
suggestions. The work of S.~P.~K. was supported by the Korea Science
and Engineering Foundation (KOSEF) grant funded by the Korea
government (MOST) (No. R01-2005-000-10404-0).


\section*{References}
\begin{thebibliography}{99}
\bibitem{Spergel} Hinshaw G  {\it et al}, 2006  Preprint [astro-ph/0603451]
\bibitem{col}     Collins C~B, Hawking~S~W, 1973
                      \textsl{Mon.\ Not.\ Roy.\ Astron.\ Soc.} {\bf 162} 307 ;
                      Barrow J~D, Juszkiewicz  R and Sonoda D~H,
                      1985
                     \textsl{ Mon.\ Not.\ Roy.\ Astron.\ Soc.} {\bf 213} 917

\bibitem{Bunn}       Bunn E~F, Ferreira  P and Silk J, 1996
                      \textsl{Phys.\ Rev.\ Lett.} {\bf 77} 2883

\bibitem{cor}        Cornish N~J{\it et al}, 2004
                      \textsl{Phys.\ Rev.\ Lett.} {\bf 92} 201302 ;\\
                      Roukema B~F{\it et al}, 2004
                     \textsl{ Astron.\ Astrophys.} {\bf 423}  821 ;\\
                      Jaffe T~R~ {\it et al}, 2005
                      \textsl{Astrophys.\ J.} {\bf 629} L1
                      [astro-ph/0606046];\\
                      Cresswell J G {\it et al}, 2006
                     \textsl{ Phys.\ Rev.\ D} {\bf 73} 041302;\\
                      Ghosh T, Hajian A  and  Souradeep T, 2006 Preprint [astro-ph/0604279]
\bibitem{wmap}   Tegmark M, de Oliveira-Costa  A ~and Hamilton A, 2003 \textsl{Phys.\ Rev.\ D}
                {\bf 68} 123523;\\
                 de Oliveira-Costa  A, Tegmark M,
                 Zaldarriage M and Hamilton A, 2004 \textsl{Phys.\ Rev.\ D} {\bf 69} 063516; \\
                  Copi C J,  Huterer D and  Starkman G D, 2003
                 Preprint [astro-ph/0310511]
\bibitem{bere}        Berera A, Buniy R~V and Kephart T~W, 2004
                     \textsl{ J. Cosmol. Astropart. Phys.}
                      JCAP 016(2004) 0410;
                      Buniy R~V, Berera A and Kephart T~W, 2006
                      \textsl{Phys.\ Rev.\ D} {\bf 73} 063529
\bibitem{campanelli} Campanelli ~L, Cea ~P and  Tedesco~L, 2006
                  \textsl{ Phys. Rev.\ Lett.}
                    {\bf97}  131302
\bibitem{deff}   Deffayet~C,~Dvali  G and Gabadadze ~G, 2002 \textsl{Phys. Rev. D}~{\bf
                 65} 044023;\\
                  Deffayet~C, Landau~S ~J,~~Raux J, Zaldarriaga M
                  and~Astier P, 2002 \textsl{Phys. Rev. D }{\bf 66} 024019 [astro-ph/0201164]
                  \bibitem{def}     Deffayet~C, 2001 \textsl{Phys. Lett. B} {\bf 502} 199
\bibitem{bin}     ~Bin$\acute{e}$truy P,Deffayet~C,~Ellwanger U and~Langlois
                D, 2000
               \textsl{Phys. Lett. B} {\bf 477} 285 [
               hep-th/9910219]
\bibitem{Taub}    Taub A H, 1951 Annals Math. {\bf 53}  472
\bibitem{sah}   ~Sahni V and ~Shtanov Y, 2003  \textsl{J. Cosmol. Astropart. Phys.} JCAP014 {
0311};\\
                R.G. Cai, Y. Gong and B. Wang, 2006 \textsl{J. Cosmol. Astropart. Phys.} JCAP006
                {0603} [hep-th/0511301]
\bibitem{isr}   ~Israel W, 1966 \textsl{Nuovo Cimento } {\bf B} {\bf 44} 1
\bibitem{deffe}   Deffayet~C, 2002  \textsl{Phys. Rev. D} {\bf 66} 103504;\\  Deffayet~C, 2005  \textsl{Phys. Rev. D} {\bf
                 71} 023520
\bibitem{bridg} ~Bridgman H A,~Malik K A and~Wands D, 2001
                 Preprint [astro-ph/01070245]
\bibitem{randall} ~Randall L and ~R.~Sundrum R, 1999 \textsl{Phys.\ Rev.\ Lett.} {\bf83} 3370;\\
               ~Randall L and ~R.~Sundrum R, 1999  \textsl{Phys.\ Rev.\ Lett.} {\bf 83}4690
\bibitem{dav}    Dvali  G, Gabadadze ~G and~~Porrati M, 2000 \textsl{Phys. Lett.
                   B}~{\bf485},
                  208 [hep-th/0005016];  Dvali  G and
                  Gabadadze, 2000 Preprint [hep-th/0008054]
\bibitem{ha} ~Ho$\rm{\check{r}}$ava P and  Witten E, 1996 \textsl{Nucl. Phys. B}~{\bf
              460} 506;\\ ~Ho$\rm{\check{r}}$ava P and  Witten E, 1996 \textsl{Nucl. Phys. B}~{\bf 475} 94
\bibitem{Spe}  ~Spergel D N {\it et al},  2003 
                      Astrophys.\ J.\ Suppl.\ {\bf 148} 175
\bibitem{ish}  Ishak M, Upadhye A and Spergel D N, 2005  Preprint [astro-ph/0507184]
\bibitem{amar} Amarzguioui M, Elgaroy O, Mota D F and Multamaki T,
                2005 Preprint [astro-ph/0510519]
\bibitem{sawi} Sawicki I and Carroll S M, 2005 Preprint [astro-ph.0510364]
\bibitem{gor} Gorbunov D, Koyama K and Sibiryakov S 2006 \textsl{Phys.
                      Rev.} {\bf D3} 044016 [hep-th/0512097]
\bibitem{confliction} Fairbairn M and Goobar A, 2006 \textsl{Phys. Lett. B }642 432 [astro-ph/0511029] ]
 \bibitem{as}          Astier P et al, 2006\textsl{ Astron. Astrophys.} 447 31
 \bibitem{song}   Song Y S, Sawicki I and Hu W, 2006 Preprint [astro-ph/0606286]
 \bibitem{ries}  Riess A et al 1998 \textsl{Astron. J.} 116 1009
 \bibitem{bar}         Barger V, Gao Y and Marfatia D, 2007 \textsl{Phys. Lett. B} {\bf 648} 127 [astro-ph/0611775]
   \bibitem{alam}                   Alam U and Sahni, 2006 \textsl{Phys. Rev. D} {\bf 73}
                      084024;\\
                      Nesseris S and Perivolaropoulos L, 2006 Preprint
                      [astro-ph/0612653]
\bibitem{sara}        Rydbeck S, Fairbairn M and Goobar A, 2007 \textsl{J. Cosmol. Astropart. Phys.}
                      JCAP05 (2007) 003
\bibitem{linder}  Linder E V, 2005 \textsl{Phys. Rev. D} {\bf 72}
                 043529 [astro-ph/0507263]
\bibitem{ma}           Maartens R and Majerotto E, 2006 \textsl{Phys. Rev. D} 74 023004 [astro-ph/0603353]


\end{thebibliography}
\end{document}